\documentclass[runningheads]{cl2emult}

\usepackage{makeidx}  
\usepackage{graphicx} 
\usepackage{subeqnar} 
\usepackage{multicol} 
\usepackage{cropmark} 
\usepackage{physbb}   
\makeindex            

\newcommand{\euler}[1]{{\usefont{U}{eur}{m}{n}#1}}

\newcommand{\umu}{\mbox{\euler{\char22}}}

\begin{document}
\title*{Dust Measurements in the Outer Solar System}
\toctitle{Dust Measurements in the Outer Solar System}
%
%
\titlerunning{Dust Measurements in the Outer Solar System}
%
\author{Eberhard Gr\"un\inst{1}
\and Harald Kr\"uger\inst{1}
\and Markus Landgraf\inst{1,}\inst{2}
}
\authorrunning{Eberhard Gr\"un et al.}
%
%
\institute{Max-Planck-Institut f\"ur Kernphysik, Heidelberg, Germany
\and NASA Johnson Space Center, Houston, TX, U.S.A.}

\maketitle              

\begin{abstract}
Dust measurements in the outer solar system are reviewed. Only the plasma wave instrument 
on board Voyagers 1 and 2 recorded impacts in the Edge\-worth-Kuiper belt (EKB). Pioneers
10 and 11 measured a constant dust flux of 10-micron-sized particles out to 20 AU. Dust
detectors on board Ulysses and Galileo uniquely identified micron-sized interstellar grains 
passing through the planetary system. Impacts of interstellar dust grains onto big EKB 
objects 
generate at least about a ton per second of micron-sized secondaries that are dispersed by 
Poynting-Robertson effect and Lorentz force. We conclude that impacts of interstellar 
particles are also responsible for the loss of dust grains at the inner edge of the EKB. While 
new dust measurements in the EKB are in an early planning stage, several missions (Cassini 
and STARDUST) are en route to analyze interstellar dust in much more detail. 
\end{abstract}

\section{Introduction}

The detection of an infrared excess at main sequence stars started a renewed interest in the 
outer extensions of our own solar system dust cloud. Especially, the observation of a dust disk 
around $\rm \beta$-Pictoris stimulated comparisons with the zodiacal cloud. Whereas the observed 
$\rm \beta$-Pictoris disk extends from a few 10 to several 100 AU distance from the central star, the 
zodiacal cloud has been observed only out to a few AU from the Sun \cite{hanner-et-al-1976}.
The zodiacal light photometer onboard Pioneer 10 observed zodiacal light out to about 3.3 
AU from the Sun where it vanished into the background. The infrared satellites IRAS and 
COBE observed dust outside the Earth's orbit. Asteroidal dust bands \cite{reach-1992}
where 
comprised of two components: a hot ($\rm \sim 300\, K$) component of nearby ($\rm \sim 1\, AU$) dust and 
a warm ($\rm \sim 200\, K$) component that corresponds to dust in the asteroid belt. 
Cold dust ($\rm < 100 \, K$) corresponds to interstellar dust beyond the solar system 
near the galactic plane. 

Detection of Edgeworth-Kuiper belt objects (EKOs) of up to a few 100 km diameter 
confirmed the existence of objects outside the planetary region in the distance range where the 
disk around $\rm \beta$-Pictoris has been found. Such objects had been theoretically predicted in 
order to explain the frequency of occurrence of short period comets and from models of the 
evolution of a disk of planetesimals in the outer solar system. About 30 objects have been 
found to date outside about 30 AU from the Sun \cite{jewitt-et-al-1998}. Theories predict 
an extension of kilometer-sized 
objects out to 3,000 AU from the Sun. Because of mutual collisions and because of their 
interactions with the environment, generation of dust has been predicted \cite{yamamoto-and-mukai-1998}
in the Edgeworth- Kuiper belt (EKB).

It is obviously very difficult to recognize faint extensions of a dust cloud when the observer 
sits inside the dense parts of this cloud, therefore, previous attempts to detect these portions of 
the zodiacal cloud by astronomical means failed. In the next section we review in-situ 
spacecraft measurements that pertain to dust in the outer solar system. Ironically, so far, the 
best evidence for dust in the outer solar system comes from measurements inside Jupiter's 
distance. 

In section~\ref{sec-3} we review our knowledge about interstellar dust in the local interstellar 
medium. After a discussion of dynamical effects of and consequences on dust in the EKB 
(section~\ref{sec-4}) we conclude in section~\ref{sec-5} by a review of future attempts and plans to get more 
information about dust in the outer solar system.

\section{Spacecraft Observations}

\begin{figure}[b]
\begin{center}
\includegraphics[width=.48\textwidth]{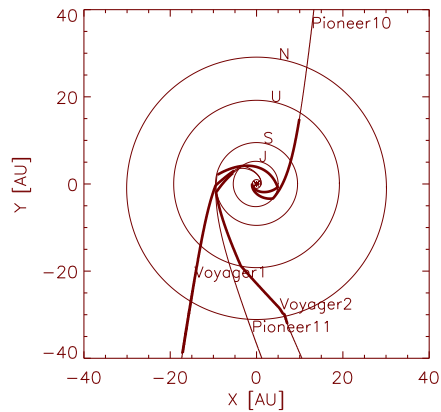}
\end{center}
\caption[]{Dust measurements in the outer solar system. Trajectories of
space spacecraft beyond Jupiter's orbit are shown in projection onto the
ecliptic plane. Heavy lines indicate regions where dust measurements
have been reported by the investigators \cite{humes-1980} 
\cite{gurnett-et-al-1997}
}
\label{fig-1}
\end{figure}
Four spacecraft carried dust detectors beyond the asteroid belt: the early Pioneers 10 and 11 
\cite{humes-1980} and recently the Galileo and Ulysses \cite{gruen-et-al-1993}
spaceprobes. Figure~\ref{fig-1} shows trajectories of spacecraft 
beyond Jupiter's orbit. The two Voyager spacecraft did not carry specific dust detectors. However, 
during the passage of Voyager 2 through the newly discovered G-ring in Saturn's ring system 
it was recognized that the plasma wave instrument onboard was able to detect impacts of 
micron-sized dust onto the spacecraft skin \cite{gurnett-et-al-1983}.
Flybys of Uranus and Neptune confirmed this effect. 
The investigators attempted to identify
dust impacts in the data obtained during occasional tracking of the spacecraft
in interplanetary space beyond 6 AU \cite{gurnett-et-al-1997}.
They found significant impact rates out to about 30 and 50 AU, respectively. The authors 
state a flux of $\rm 10^{-11}\,g$ particles of about $\rm 5\cdot 10^{-4}\,m^{-2}\,s^{-1}$. 
This flux is more than a factor 10 above the corresponding zodiacal dust flux at 1 AU 
\cite{gruen-et-al-1985}.
The problem with the chance dust detector onboard the Voyager spacecraft is (1) 
that this instrument has never been calibrated for dust detection, i.e. its sensitivity 
has not been experimentally 
determined, (2) the sensitive area of the detector has only been derived from theoretical 
considerations, and (3) the distinction of impact events from noise events has not been 
verified, i.e. observations of micron-sized dust by Voyager have only been reported from 
regions of space where no other spacecraft took similar measurements. Only Cassini 
measurements may confirm the Voyager findings and provide a cross calibration. Therefore, 
quantitative results from the Voyager observations have to be taken with great caution.

Beginning at about 3 AU from the Sun, measurements of about 10 micron-sized dust 
by the Pioneer 10 detector (mass threshold $\rm \sim 8 \cdot 10^{-9}\, g$) 
showed a constant dust density out to 18 AU \cite{humes-1980}.
At this distance the nitrogen in the pressurized dust sensor froze 
out and prohibited measurements further away from the Sun. Dust measurements 
by Pioneer 11 (mass threshold $ \rm \sim~6~\cdot~10^{-9}\, g$) 
out to Saturn's distance were reported by Humes \cite{humes-1980}.
During the three passages of Pioneer 11 through the heliocentric distance 
range from 3.7 to 5 AU the detector observed a roughly constant dust flux. This led Humes to 
conclude that this dust had to be on highly eccentric orbits that have random inclinations (if 
the particles are on bound orbits about the Sun). 

Interstellar dust grains passing through the planetary system have been detected by the dust 
detector onboard the Ulysses spacecraft \cite{gruen-et-al-1993}.
These observations provided the
unique identification of interstellar grains by three characteristics: 1. At Jupiter's distance the 
grains seemed to move on retrograde trajectories opposite to orbits of most 
interplanetary grains and the flow direction coincided with that of interstellar gas 
\cite{witte-et-al-1993},
2. A constant flux has been observed at all latitudes above the ecliptic plane, while 
interplanetary dust displays a strong concentration towards the ecliptic
\cite{gruen-et-al-1997} \cite{krueger-et-al-1998} and 3. The measured 
speeds (despite their substantial uncertainties) of the interstellar grains were high 
($\rm \geq 15\,km\,s^{-1}$) which 
indicated orbits unbound to the solar system, even if one neglects radiation pressure effects
\cite{gruen-et-al-1994}. 

\section{Interstellar Dust Characteristics}  \label{sec-3}

\begin{figure}
\begin{center}
\includegraphics[width=.49\textwidth]{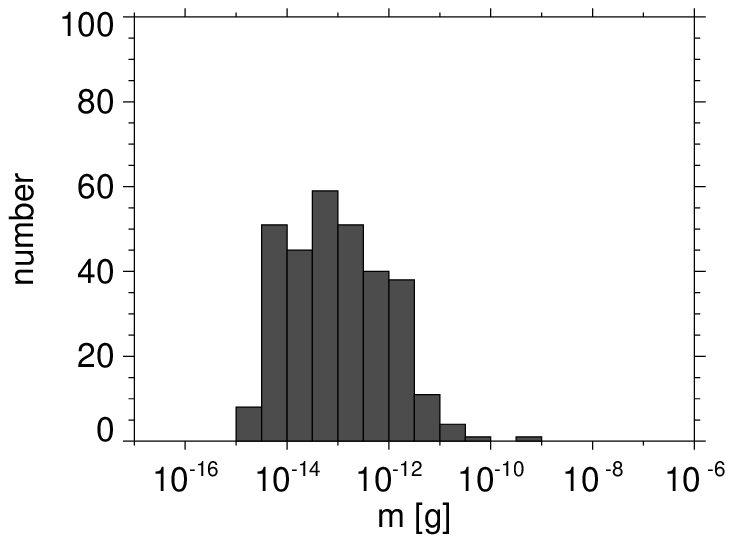}
\includegraphics[width=.49\textwidth]{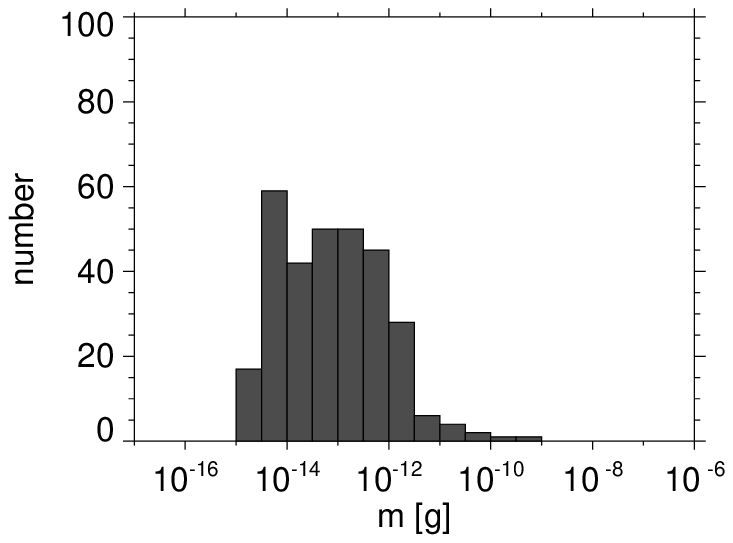}
\end{center}
\caption[]{
Mass distributions of interstellar grains observed by the Galileo (left)
and Ulysses (right) dust instruments \cite{landgraf-1998}.
The detection
threshold of the detectors is $\rm 10^{-15}\, g$ at $\rm 26\,km\,s^{-1}$
}
\label{fig-2}
\end{figure}
Clearly identified interstellar grains range from $\rm 10^{-15}\, g$ to above $\rm 10^{-11}\, g$ 
(see Figure~\ref{fig-2}) with a maximum at about $\rm 10^{-13}\, g$. The deficiency of small grain 
masses is not solely introduced by the detection threshold of the instrument but indicates a 
depletion of small interstellar grains in the heliosphere. Estimates of the filtering of 0.1 
micron-sized and smaller electrically charged grains in the heliospheric bow shock region 
\cite{frisch-et-al-1999}
and in the heliosphere itself \cite{landgraf-1998}
show that these small particles 
are strongly impeded from entering the planetary system by the interaction with the ambient 
magnetic field.

\begin{figure}
\begin{center}
\vspace{-2.7cm}
\includegraphics[width=1.0\textwidth]{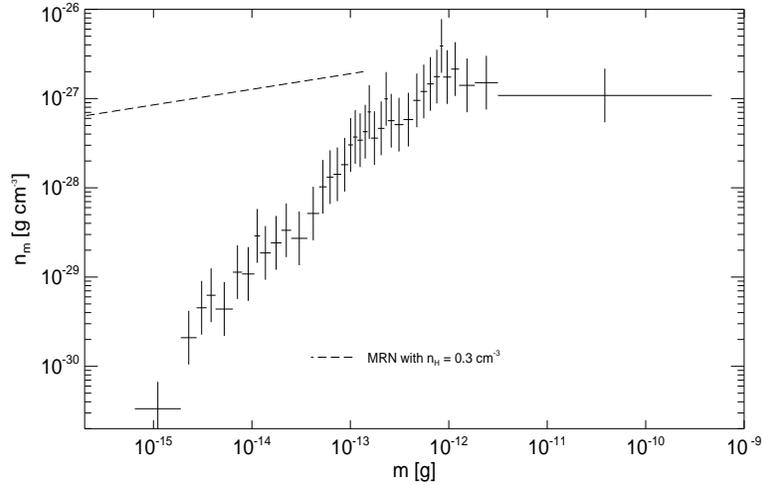}
\vspace{-5.3cm}
\end{center}
\caption[]{
Mass density of interstellar dust. Measurements by Galileo and
Ulysses in the inner heliosphere are compared with "classic"
astronomical grains expected to be present in the local interstellar
medium as well. The astronomical grains are represented by the MRN
distribution \cite{mathis-et-al-1977}
corresponding to a total hydrogen density of $\rm 0.3\, cm^{-3}$
}
\label{fig-3}
\end{figure}

The mass density of interstellar grains detected by Galileo and Ulysses is displayed in 
Figure ~\ref{fig-3}. Below about $\rm 10^{-13}\, g$ it shows a strong deficiency of small grains 
due to 
heliospheric filtering. Above about $\rm 10^{-12}\, g$ a flat distribution is suggested 
(corresponding to a slope of -4 of a differential size distribution). The total mass density of the 
observed grains is $\rm 7\cdot  10^{-27}\, g\, cm^{-3}$. The upper limit ($\rm 10^{-9}\, g$) of the 
mass distribution is not well determined: if we extend the flat distribution to bigger masses,
about $\rm 1.5\cdot 10^{-27}\, g\, cm^{-3}$ per logarithmic mass interval has to be added. 

Even bigger radar meteor particles ($\rm m \ge 10^{-7} g$) have been found \cite{taylor-et-al-1996} 
to enter the Earth's atmosphere with speeds above $\rm 100 \, km\,s^{-1}$. These speeds are well 
above the escape speed from the solar system which confirms their interstellar origin. Big 
interstellar meteors arrive from a broad range of directions and are not collimated to the 
interstellar gas direction as smaller particles are. At present the total mass flux of big 
interstellar meteor particles is not known. Therefore, an extrapolation from the Ulysses 
observations up to $\rm 10^{-7} g$ has a large uncertainty. 

Frisch et al. \cite{frisch-et-al-1999} summarize properties of the local interstellar cloud 
(LIC, Table~\ref{tab-1}). If the 
total hydrogen density is complemented by helium (with a number density ratio $\rm n_{He}/n_{H} = 0.1$) 
the total gas mass density in the LIC is about $\rm 7\cdot  10^{-25}\, g\, cm^{-3}$. The canonical 
gas-to-dust mass ratio of 100 (from "cosmic abundance" considerations) compares favorable with 
the observed values. However, several modifications of the dust mass density in the LIC are 
suggested. Firstly, small "classic" astronomical interstellar grains may need to be added
to the interstellar grains detected by Galileo and Ulysses. 
Mathis, Rumpel and Nordsieck (MRN, \cite{mathis-et-al-1977})
represent these particles by a power law with exponent -3.5 in the radius range from 
5 to 250 nm. Figure~\ref{fig-3} shows this distribution by a dashed
line. A mass density of 5 to $\rm 7\cdot 10^{-27}\, g\, cm^{-3}$ has to be added if these small 
particles are present in the LIC - which we will assume in the further discussion. Secondly, 
supernova shocks passing through the interstellar medium process interstellar grains by 
shattering and evaporation, i.e. part of the grain material is put into the gas phase and shows 
up as absorption lines in the spectra of nearby stars. Refractory elements like Mg, Si, Ca, and 
Fe have been identified. Therefore, the total content of heavy elements in the LIC is further 
increased. 
Both the mere existence of big interstellar grains ($\rm > 10^{-13}\, g$) and the total mass of 
interstellar grains in the LIC have important consequences for the understanding of the 
interstellar medium. Big grains couple to the interstellar gas over much longer lengths scales 
than the small "classic" interstellar grains, both by friction and by gyro-motion imposed 
by the  interstellar magnetic field. Grains in the diffuse interstellar medium are electrically 
charged by the competing effects of electron collection from the ambient medium and the 
photo-effect of the far UV radiation field. The so charged dust grains couple to the magnetic 
field which itself is strongly coupled to the ionized component of the interstellar medium. A 
simple comparison shows that the electromagnetic coupling length is several orders of 
magnitude shorter than the frictional length scale for LIC conditions. However, for particles 
with masses $\rm > 10^{-12}\, g$ the gyro radius exceeds the dimension of the LIC and, 
therefore, these particles are not expected to move with the LIC gas. $\rm 10^{-7} \, g$ particles 
could travel more than 100 pc through the diffuse interstellar medium (at LIC conditions) with 
little effect. This mechanism provides the basis for any heterogeneity in the gas-to-dust mass 
ratio. Locally there may be significant variations in the gas-to-dust mass ratio and hence 
deviations from the "cosmic abundance" which has to be preserved only on the average over 
large regions of space. 
 
\begin{table}[t]
\caption{\label{tab-1}
Characteristics of the local interstellar cloud (LIC, after \cite{frisch-et-al-1999}).
Both hydrogen number densities (neutral: $\rm n_{H0}$, and ionized $\rm n_{H+}$) are given}
\begin{center}
\renewcommand{\arraystretch}{1.4}
\setlength\tabcolsep{5pt}
\begin{tabular}{cc}
\noalign{\smallskip}
\hline
Item&Value\\
\hline\noalign{\smallskip}
$\rm n_{H0}$     &  $\rm 0.22\, cm^{-3}$\\
$\rm n_{H+}$     &  $\rm 0.1\, cm^{-3}$\\
Temperature      &      6,900\, K\\
Magnetic Field   &   0.15 to 0.6\, nT\\
\hline
\end{tabular}
\end{center}
\label{Tab1a}
\end{table}

\section{Dust Dynamics in the Edgeworth-Kuiper Belt}  \label{sec-4}

Impacts of interstellar grains onto objects in the Edgeworth-Kuiper belt generate dust locally. 
In order to estimate the amount of dust generated we represent the size distribution of EKOs 
by a simple power law \cite{stern-1996} \cite{jewitt-et-al-1998} 
$\rm n(s)\, ds = N_0\, s^{-4}\, ds$, 
with $\rm N_0 = 1.3\cdot 10^{19}$, i.e. 35,000 objects are in the observed EKO size range of 
$\rm 5\cdot 10{^4}\,m$ to $\rm s_{max} = 1.6\cdot 10^{5}\, m$ radius. This distribution has constant 
mass per equal logarithmic size interval, 
but most of the cross sectional area is in the smallest objects. Therefore, the size distribution is 
truncated at $\rm s_{min} = 100\, m$. By integrating this size distribution over the full size range we 
arrive at a total cross section of about $\rm 4\cdot 10^{17}\, m^{2}$. 

Impact experiments with micron-sized projectiles into water ice \cite{eichhorn-and-gruen-1993}
suggest that at $\rm 26\, km\,s^{-1}$ impact speed about $\rm 10^{4}$ times the projectile mass is 
excavated and ejected mostly in form of small particulates. Because of the low gravity of 
EKOs we assume that most secondary particles are ejected at speeds in excess of the escape 
speed of 0.1 to $\rm 10\, m\, s^{-1}$. The interstellar dust mass flux of 
$\rm 2\cdot 10^{-16}\, g\, m^{-2}\, s^{-1}$ 
generates ejecta particles at a rate of $\rm 8\cdot 10^{5}\, g\,s^{-1}$. A more detailed calculation 
\cite{yamamoto-and-mukai-1998} arrives at similar values. About the same amount of micron-sized 
dust is generated by mutual collisions among EKOs \cite{stern-1996},
i.e. about 2 tons of dust are 
generated in the Edgeworth-Kuiper belt every second. This value has a large uncertainty and is 
probably a lower limit since much cross sectional area could be in smaller EKOs than 
previously assumed. 

In the absence of big planets the Poynting-Robertson effect is the most important dynamical 
effect on dust in the EKB. The time $\rm \tau_{pr}$ (years) for a dust grain of radius s (cm) and density 
$\rm \rho_d (g\, cm^{-3}$) to spiral to the Sun from a circular orbit at distance r (AU) under the Poynting-
Robertson effect can be estimated from 
\begin{equation}
   \rm        \tau_{pr} = \rm 7\cdot 10^{6}\, s\, \frac{\rho_d}{Q_{pr}} r^2 .   \label{eqn-1}
\end{equation}
A 10 \umu m sized particle with $\rm \rho_d = 2.7\, g\,cm^{-3}$ and radiation pressure 
efficiency $\rm Q_{pr} = 1$ would need 
about 6.5 Myrs to reach the inner planetary system starting from an initial circular orbit at 
50 AU. The radiation pressure efficiency $\rm Q_{pr}$ decreases for particles smaller than the effective 
wavelength of solar radiation. Liou et al. 
\cite{liou-et-al-1996} have shown that during their orbital evolution 
micron-sized grains are trapped in mean motion resonances with the outer giant planets. The 
biggest effect comes from resonances with Neptune which prolongs the particle's residence 
time inside about 40 AU significantly. Therefore, the Poynting-Robertson life time given in 
eqn.~(\ref{eqn-1}) is only a lower limit for the dynamical life time of EKB dust particles. Liou et al. found 
that in many cases ($\rm \sim 80\,\%$) particles are ejected out of the solar system by passages close to 
Jupiter before they could reach the region of the inner planets. 

There is another force that becomes increasingly important in the outer heliosphere: Lorentz 
scattering of electrically charged interplanetary grains by solar wind magnetic field 
fluctuations \cite{morfill-et-al-1986}.
Charging of dust particles by the combined solar UV 
photo-effect and electron capture from solar wind plasma results in a surface potential of about 
+5 V leading to a charge-to-mass ratio that varies as $\rm s^{-2}$. Carried out by the solar 
wind plasma (at speeds of 400 to $\rm 800\, km\,s^{-1}$ away from the Sun) the interplanetary magnetic 
field forms a Parker spiral. The dominant azimuthal component of the magnetic field varies as 
1/r with heliocentric distance. At 1 AU the Lorentz force is comparable to solar gravitational 
attraction for particles of $\rm s \sim 0.1$ microns. Therefore, in the EKB at 50 AU, Kepler orbits of 
even micron-sized particles are strongly affected by the Lorentz force. 

Near the solar equatorial plane ($\rm \sim  \pm 15^{\circ}$) the magnetic field changes polarity two to 
four times per solar rotation period (25.2 d). Above and below this equatorial region (which is 
roughly centered at the ecliptic plane) a unipolar magnetic field prevails that changes its 
polarity with the 11-year solar cycle. Both short and long-term magnetic field fluctuations lead 
to diffusion of grain orbits mostly in inclination \cite{morfill-et-al-1986}
but also some outward 
convection reduces the Poynting-Robertson effect of 10 micron-sized and smaller grains. 

This dynamical evolution has to be compared with the collisional life times in the outer solar 
system. There the dominant flux of micron-sized projectiles is from interstellar grains. 
Therefore, we calculate the collision rate $\rm C_{coll}$ and the corresponding mean collisional life time 
$\rm \tau_{coll} = 1/ C_{coll}$ for interstellar grains. We follow a similar calculation for interplanetary grain 
collisions by Gr\"un et al. \cite{gruen-et-al-1985}, especially, we use the same collisional parameters for our 
calculation. Although the EKB is beyond the planetary region it is still located inside the 
heliosphere and some filtering of interstellar grains may occur. Therefore, we cut-off the 
interstellar size distribution at about $\rm 10^{-15} g$. 

\begin{figure}[h]
\begin{center}
\includegraphics[width=.7\textwidth]{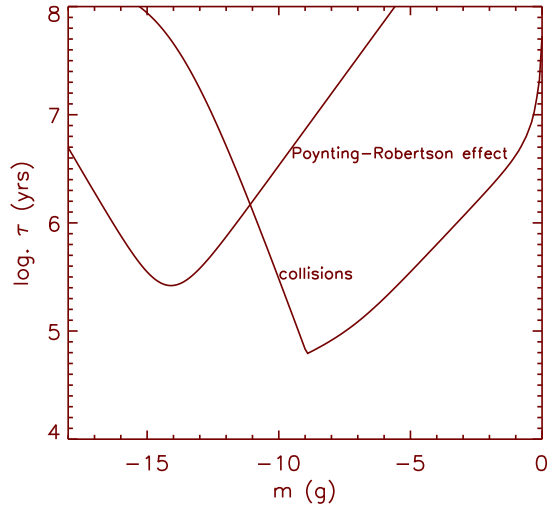}
\end{center}
\caption[]{Life times of dust grains at 50 AU from the Sun. Collision
life times are calculated for a flux of interstellar projectiles with
masses $\rm > 10^{-15}\, g$. For comparison life times due to the Poyting-Robertson
effect are shown. The Poyting-Robertson life time is calculated for
particles on circular orbits starting at 50 AU distance; no planetary
resonance effects have been considered
}
\label{fig-4}
\end{figure}

Figure~\ref{fig-4} shows the life times of EKB particles at 50 AU due to collisions with 
interstellar grains. For comparison we show the pure Poynting-Robertson life times. 
The life times of 10 micron-sized particles are dominated by collisions with 
interstellar grains. Considering the prolonged residence time due to Lorentz scattering and 
mean motion resonances, even smaller EKB grains are destroyed by interstellar dust impacts 
before they can reach the inner planetary system where they mix-in with zodiacal dust. We 
conclude that impacts of interstellar particles are not only a major contributor of dust in the 
EKB but may also be responsible for the loss of dust grains at the inner edge of the EKB. A 
complication is that the density of the ambient interstellar medium is variable on time scales 
of $\rm 10^{5}$ to $\rm 10^{6}$ years and extrapolations from the present state cannot be easily 
made. Impacts of interstellar grains may play an important role for the existence and structure 
of extended dust sheets like that around $\rm \beta$-Pictoris.

\section{Future Measurements}  \label{sec-5}

New measurements of interstellar grains passing through the planetary system are expected 
from the Cassini and STARDUST missions. Cassini with its Cosmic Dust Analyzer (CDA) 
was launched in October 1997. It will commence dust measurements at its final Venus 
flyby in June 1999 and continue to make interplanetary dust measurements until its arrival at 
Saturn in 2004. The Cassini CDA combines a large area dust detector ($\rm 0.1\,m^2$) 
with a mass analyzer for 
impact generated ions. Thereby, the first medium-resolution ($\rm M/ \Delta M \sim 20$ to 50) 
compositional measurements of interstellar grains will be performed. 

The STARDUST Discovery mission will collect samples of cometary coma and interstellar 
dust and return them to Earth. Several times during its eccentric orbit about the Sun (out to 
about 3 AU) interstellar dust in addition to dust from Comet Wild 2 will be captured by 
impact into aerogel and brought back to the Earth in 2006. In addition in-situ detection and 
high-resolution ($\rm M/ \Delta M > 100$) compositional measurements of cometary and interstellar 
grains will be performed by the Cometary and Interstellar Dust Analyzer (CIDA). Interstellar 
dust analyses and collections may be possible even in high-Earth orbit (Gr\"un, in 
preparation). 

The currently studied Pluto Kuiper Express mission focuses on the big objects Pluto and 
Charon and perhaps one EKO, but no dust measurements in the EKB are considered. Missions 
to the heliospheric boundary and beyond are in their early planning phases and have to 
take into account dust in the outer solar system at least for hazard studies.

\clearpage
\addcontentsline{toc}{section}{Index}
\flushbottom
\printindex

\end{document}